\documentclass[a4paper,fleqn,usenatbib]{mnras}

\usepackage[T1]{fontenc}
\usepackage{ae,aecompl}

\usepackage{graphicx}	% Including figure files
\usepackage{amsmath}	% Advanced maths commands
\usepackage{amssymb}	% Extra maths symbols

\title[$\rm B_{pos}$ and dust filaments in the Polaris Flare]{The magnetic field and dust filaments in the Polaris Flare}

\author[G. V. Panopoulou et al.]{
G. V. Panopoulou,$^{1,2}$\thanks{E-mail: panopg@physics.uoc.gr}
I. Psaradaki,$^{1}$
K. Tassis$^{1,2}$
\\
$^{1}$Department of Physics and ITCP\thanks{Institute for
  Theoretical and Computational Physics, formerly Institute for Plasma
Physics}, University of Crete, 71003, Heraklion, Greece\\
$^{2}$Foundation for Research and Technology - Hellas, IESL, Voutes, 71110 Heraklion, Greece\\
}

\date{Accepted XXX. Received YYY; in original form ZZZ}

\pubyear{2016}

\begin{document}
\label{firstpage}
\pagerange{\pageref{firstpage}--\pageref{lastpage}}
\maketitle

\begin{abstract}
In diffuse molecular clouds, possible precursors of star-forming clouds, the effect of the magnetic field is 
unclear. In this work we compare the orientations of filamentary structures in the Polaris Flare, as seen through 
dust emission by \textit{Herschel}, to the plane-of-the-sky magnetic field orientation ($\rm B_{pos}$) as revealed 
by stellar optical polarimetry with RoboPol. Dust structures in this translucent cloud show a strong preference for 
alignment with $\rm B_{pos}$. 70\% of field orientations are consistent with those of the filaments
(within 30$^\circ$). 
We explore the spatial variation of the relative orientations and find it to be uncorrelated with the 
dust emission intensity and correlated to the dispersion of polarization angles. 
Concentrating in the area around the highest column density filament, and in the region 
with the most uniform field, we infer the $\rm B_{pos}$ strength to be 24 $-$ 120 $\mu$G. 
Assuming that the magnetic field can be decomposed into a turbulent and 
an ordered component, we find a turbulent-to-ordered ratio of 0.2 $-$ 0.8, implying that the magnetic 
field is dynamically important, at least in these two areas. We discuss implications on the 3D field 
properties, as well as on the distance estimate of the cloud.
\end{abstract}

\begin{keywords}
ISM: magnetic fields -- ISM: structure -- ISM: clouds -- ISM: individual objects: Polaris Flare -- polarization 
\end{keywords}

%%%%%%%%%%%%%%%%%%%%%%%%%%%%%%%%%%%%%%%%%%%%%%%%%%

%%%%%%%%%%%%%%%%% BODY OF PAPER %%%%%%%%%%%%%%%%%%

\section{Introduction}
The structure of interstellar clouds is highly complex, characterized by the existence of elongations, or
filaments, of various scales \citep[e.g.][]{myers2009}. The Gould Belt Survey conducted by the \textit{Herschel}
space observatory captured the morphologies of the nearby molecular clouds with unprecedented sensitivity and 
detail, allowing the study of filamentary structures to advance \citep[e.g.][]{andre2010}.  
A better understanding of filament properties and their relation to their environments could provide
clues as to how clouds proceed to form stars. 

To this end, important questions to answer are whether the magnetic field interacts with filaments and how this 
interaction takes place. Its role in the various stages and environments of star formation is hotly debated. 
In simulations of super-Alfv\'{e}nic turbulence, magnetic fields are tangled due to the gas flow 
\citep[e.g.][]{ostriker, falceta}. In such models, filaments are formed by shock interactions 
\citep[][]{heitsch2001, padoan2001} and the magnetic field within them is found to lie along their spines 
\citep{heitsch2001, ostriker, falceta}. In studies of sub/trans-Alfv\'{e}nic turbulence, where the magnetic field is 
dynamically important, filament orientations with respect to the large scale ordered field, depend on whether 
gravity is important. In simulations where gravity is not taken into account \citep[e.g.][]{falceta} or structures 
are gravitationally unbound \citep[][]{soler2013}, filaments are parallel to the magnetic field. 
On the other hand, self-gravitating elongated structures are perpendicular to the magnetic field
\citep{mouschovias1976, nakamura2008, soler2013}. Both configurations are the result of the magnetic force
facilitating flows along field lines.
Finally, if the magnetic field surrounding a filament has a helical configuration \citep{fiege} the 
relative orientation of the two as projected on the plane of the sky can have any value, depending on 
projection, curvature and if the field is mostly poloidal or toroidal \citep{matthews2001}.

The relation between cloud structure and the magnetic field was highlighted early on by 
observational studies \citep[e.g. in the Pleiades,][]{hall1955}.
On cloud scales, \cite{li2013} found that the distribution of relative orientations of 
elongated clouds and the 
magnetic field (both projected on the plane of the sky) is bimodal: clouds either tend to be parallel or 
perpendicular (in projection) to the magnetic field.
In a series of papers the \textit{Planck} collaboration compared the magnetic field to
ISM structure across a range of hydrogen column densities ($\rm N_H$). \cite{planck32} considered the orientation 
of structures in the diffuse ISM in the range of $\rm N_H\sim 10^{20} - 10^{22}$ and found significant alignment 
with the plane-of-the-sky magnetic field ($\rm B_{pos}$). \cite{planck35} found that in their sample of 10 nearby 
clouds, substructure at high column density tends to be perpendicular to the magnetic field, 
whereas at low column density there is a tendency for alignment.

Studies of optical and NIR polarization, tracing $\rm B_{pos}$ in cloud envelopes, show that dense filaments 
within star-forming molecular clouds tend to be perpendicular to the magnetic field \citep[][]{pereyra2004, 
alves2008, chapman2011, sugitani2011}. On the other hand, there are diffuse linear structures termed 
\textit{striations}, that share a common smoothly varying orientation and are situated either in the outskirts of 
clouds \citep{goldsmith2008, deoliveira} or near dense filaments \citep{palmeirim}. 
These structures show alignment with $\rm B_{pos}$ \citep[][]{vandenbergh, chapman2011, palmeirim, deoliveira,  malinen2015}. 
The extremely well-sampled data of \cite{franco2015} in Lupus I show that $\rm B_{pos}$ is perpendicular to 
the cloud's main axis but parallel to neighbouring diffuse gas. There are, however, exceptions to this trend 
\citep[e.g. L1506 in Taurus,][]{goodman1990}.

While the large-scale magnetic field has been mapped in many dense molecular clouds, little is known about the
field in translucent molecular clouds. In this paper we investigate the relation of the magnetic field to the gas 
in the Polaris Flare, a high-latitude diffuse cloud. This molecular cloud extends above the galactic plane and 
is at an estimated distance between 140 and 240 pc, although this is debated 
\citep[e.g.][]{zagury1999, schlafly2014}. It is a translucent region \citep[$\rm A_V \lesssim 1$ mag,][]{cambresy} 
devoid of star formation activity \citep{andre2010, menshchikov, WT2010}, except for the existence of 
possibly prestellar core(s) in the densest part of the cloud MCLD 123.5+24.9 (MCLD123) 
\citep[][]{WT2010,  wagle}. Signatures of intense velocity shear have been identified in this region and have been 
linked to the dissipation of supersonic (but trans-Alfv\'enic) turbulence 
\citep[][and references therein]{hily-blant2009}.

The dust emission in $\sim$16 deg$^2$ of the Polaris Flare has been mapped by \textit{Herschel}
\citep{pilbratt2010} as part of the \textit{Herschel} Gould Belt Survey 
\citep{andre2010,miville,menshchikov,WT2010}. The \textit{Planck} space observatory has provided the first map 
of the plane-of-the-sky magnetic field in the area at tens of arcminute resolution \citep{planck19, planck20}. 
In \cite{panopoulou2015} (Paper I) we presented a map of the plane-of-the-sky magnetic field in the same area, 
measured by stellar optical polarimetry with RoboPol. The resolution of optical polarimetry (pencil beams) and 
coverage of our data allow for a detailed comparison between magnetic field and cloud structure. 
The goal of this work is to compare the magnetic field orientation to that of the linear structures in the 
Polaris Flare and to estimate the plane-of-the-sky component of the field in various regions of the 
cloud. In section \ref{ssec:wholemap} we present the distribution of relative orientations of filaments and $\rm 
B_{pos}$ throughout the mapped region. We compare properties such as the relative orientations and polarization 
angle dispersion across the map in section \ref{ssec:maps} and present the distribution of filament widths in 
section \ref{ssec:width}. We analyse two regions of interest separately in section \ref{ssec:regions} and 
estimate the $\rm B_{pos}$ strength in these regions in section \ref{ssec:Bstrength}. 
Finally, we discuss implications of our findings in section \ref{sec:discussion} and summarize our results in 
section \ref{sec:summary}.

\section{Results} 
\label{sec:results}
\subsection{Relative orientations of $\rm B_{pos}$ and dust filaments}
\label{ssec:wholemap}

\begin{figure*}
\centering
\includegraphics[scale=1]{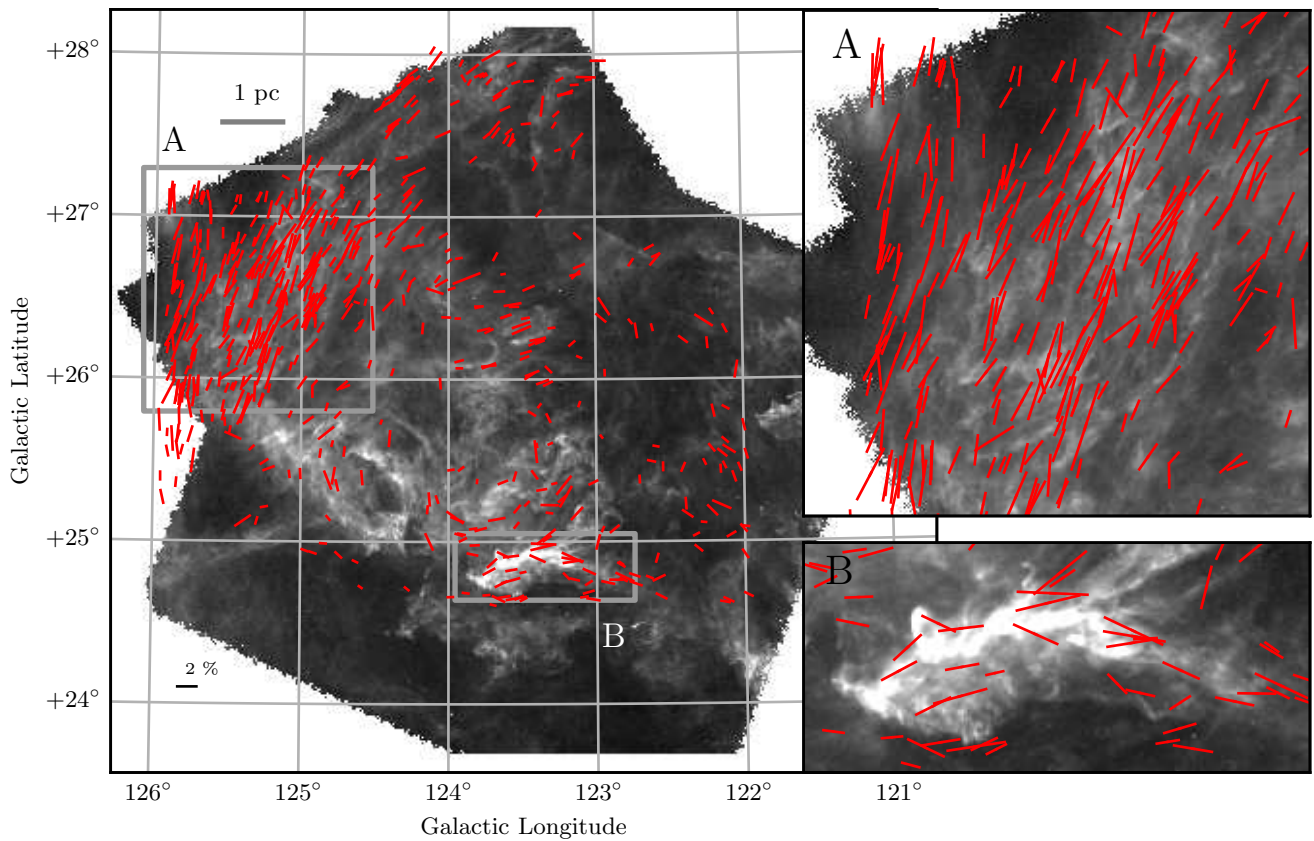}
\caption{\textit{Herschel} 250 $\rm\mu m$ image of the Polaris Flare (grayscale). Segments are the optical
polarization data from Paper I. Regions A (striations) and B (MCLD123), indicated by the rectangles in
the full map, are shown in the top right and bottom right inlets for detail.
Scales of 1 pc and $p_d = $ 2\% are shown in the top left and bottom left corners of the full map.}
\label{fig:regions}
\end{figure*}

We use the \textit{Herschel}-SPIRE 250 $\rm\mu m$ map\footnote{The map is part of the preliminary data available at 
the \textit{Herschel} Gould Belt Survey archive.} in our analysis of cloud structure.
Filamentary structures are evident throughout the \textit{Herschel} image of this translucent cloud.
We will compare the orientations of these structures to the optical polarization data presented\footnote{Following
the publication of the catalogue of 609 measurements in Paper I, 
we discovered an error in the conversion from polarization angle with respect to the North Celestial 
Pole to angle with respect to the North Galactic Pole. This error affected the orientation of 77 segments in the 
top of the published map. We corrected the formula used for the conversion
and have updated the values of galactic angle in the published catalogue.
This updated version is used in this work and can be found 
at http://cds.u-strasbg.fr/ No other changes have been made to the original release.} in Paper I, shown as line 
segments on the 250 $\rm\mu m$ dust emission image in 
Fig. \ref{fig:regions}. The segment length is proportional to the (debiased) fractional linear polarization, $p_d$. 
A line showing a $p_d$ of 2\% is located in the bottom left corner and a line of 1 pc length is located in the 
top left. We adopt a distance of 150 pc to be consistent with the 
analysis of \textit{Herschel} data in the literature \citep[based on][]{heithausen1999}.
The rectangles highlight regions of interest that will be referred to in the following analysis. 
Inlets on the right show zoomed-in versions of both regions.
A preliminary inspection of the map shows that the orientation of $\rm B_{pos}$ revealed from optical polarimetry 
seems to correlate well with the apparent orientations of filaments in most of the map.
This correlation can be quantified by measuring the relative orientation of the two.

An excellent tool for determining the orientation of linear structures, irrespective of their brightness, 
is the Rolling Hough Transform (RHT) \citep{clark2014}. It was introduced in the study of diffuse HI and 
has been used in analyses of molecular clouds as well \citep{koch2015,malinen2015}. 
The RHT quantifies the linearity of cloud structure for every image pixel. 
It does so by measuring the intensity along any given direction within a disk region surrounding each pixel.
The RHT returns the probability that a pixel is part of a linear structure as a function of angle.
Integrating for all angles results in a visualization of linear features in the image (RHT backprojection).

\begin{figure}
\centering
\includegraphics[scale=1]{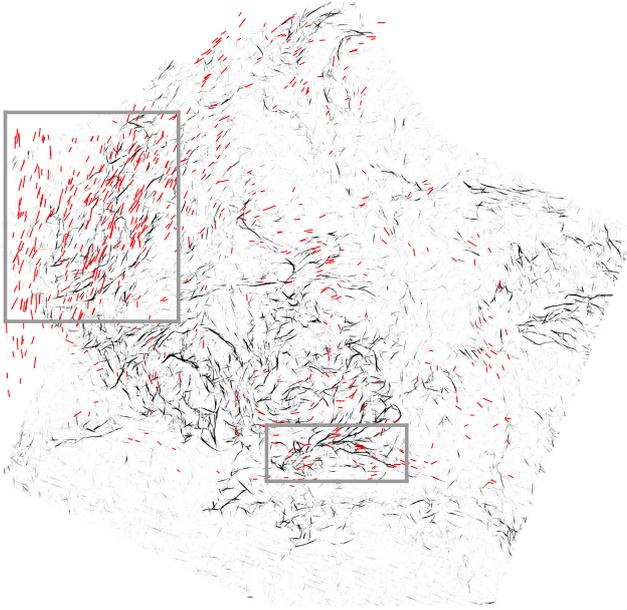}
\caption{RHT visualization of the \textit{Herschel} 250 $\rm\mu m$ image of the Polaris Flare (grayscale). 
Dark pixels correspond to a high probability of linearity. 
The optical polarization segments from Paper I have been overplotted in red. Regions A and B are outlined as in Fig. 1.}
\label{fig:rhtmap}
\end{figure}

We applied the RHT to the \textit{Herschel} 250 $\rm\mu m$ image and present the RHT backprojection in Fig. 
\ref{fig:rhtmap}. The darkest pixels in the image belong to well-defined linear structures in intensity (filaments 
and striations). Apart from these structures, the RHT backprojection contains some spurious features, e.g. in the bottom
part of the map where dust emission is very faint or absent. These features seem correlated with the \textit{Herschel}
scanning direction. Since neither cloud structure nor magnetic field orientation coincide with this direction, the effect of
these features on the distribution of relative orientations will be to randomize a very small number (if any) of values, 
introducing misalignment.

\begin{figure}
\centering
\includegraphics[scale=1]{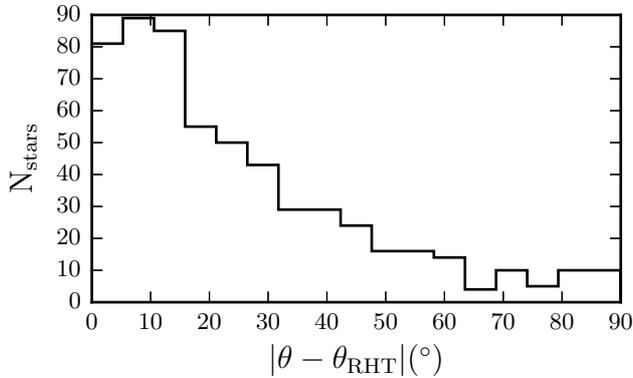}
\caption{Distribution of the absolute difference between the angle of each polarization segment and the RHT angle 
in its vicinity ($\rm \vert\theta - \theta_{RHT}\vert$). The distribution contains all 570 polarization 
measurements that lie in the \textit{Herschel} image.}
\label{fig:relativethetas}
\end{figure}

The polarization segments tracing the orientation of $\rm B_{pos}$ are overplotted on the backprojection in Fig. \ref{fig:rhtmap}. 
At the position of each polarization measurement, we calculate the mean RHT angle 
\citep[$\rm \theta_{RHT}$, defined as in][]{clark2014}, 
within a circle of diameter 0.17 pc (corresponding to an angular diameter of 4\arcmin). 
We then compare $\rm \theta_{RHT}$ to the orientation of each polarization segment, 
$\rm \theta$, by taking the absolute value of their difference\footnote{All angles are defined with respect to the 
North and increase towards the East, according to the IAU convention.}, $\rm \vert\theta - \theta_{RHT}\vert$. 
There are 39 polarization measurements that extend outside the dust emission image and are not included in the 
comparison to cloud structures. The distribution of the relative orientations for the 570 remaining 
measurements is plotted in Fig. \ref{fig:relativethetas}. 
There is a strong preference in alignment: 70\% of polarization measurements are within 30$^\circ$ of the 
orientation of linear features in their surrounding gas. A Monte-Carlo run showed that the probability of
obtaining this correlation by chance is less than $10^{-6}$.
Only 8\% are within 30$^\circ$ of being perpendicular to their surrounding gas.
We explored the effect of changing the parameters of the RHT, as well as the area around each star used in 
calculating $\rm \theta_{RHT}$, and found no significant variation for a large portion of the parameter space.

\subsection{Variation across the field}
\label{ssec:maps}
The results presented above on the alignment of filaments and $\rm B_{pos}$ concerned the entire map. 
However, both field and cloud structure are not homogeneous across the cloud. 
For example, there is a low column density area of striations, marked by the rectangle as region A in 
Fig. \ref{fig:regions}, where $\rm B_{pos}$ exhibits ordered structure. Adjacent to this area, towards 
lower latitude and longitude, in a significant portion of the map, measurements are sparse and polarization angles
show substantial dispersion. 

\begin{figure*}
\centering
\includegraphics[scale=1]{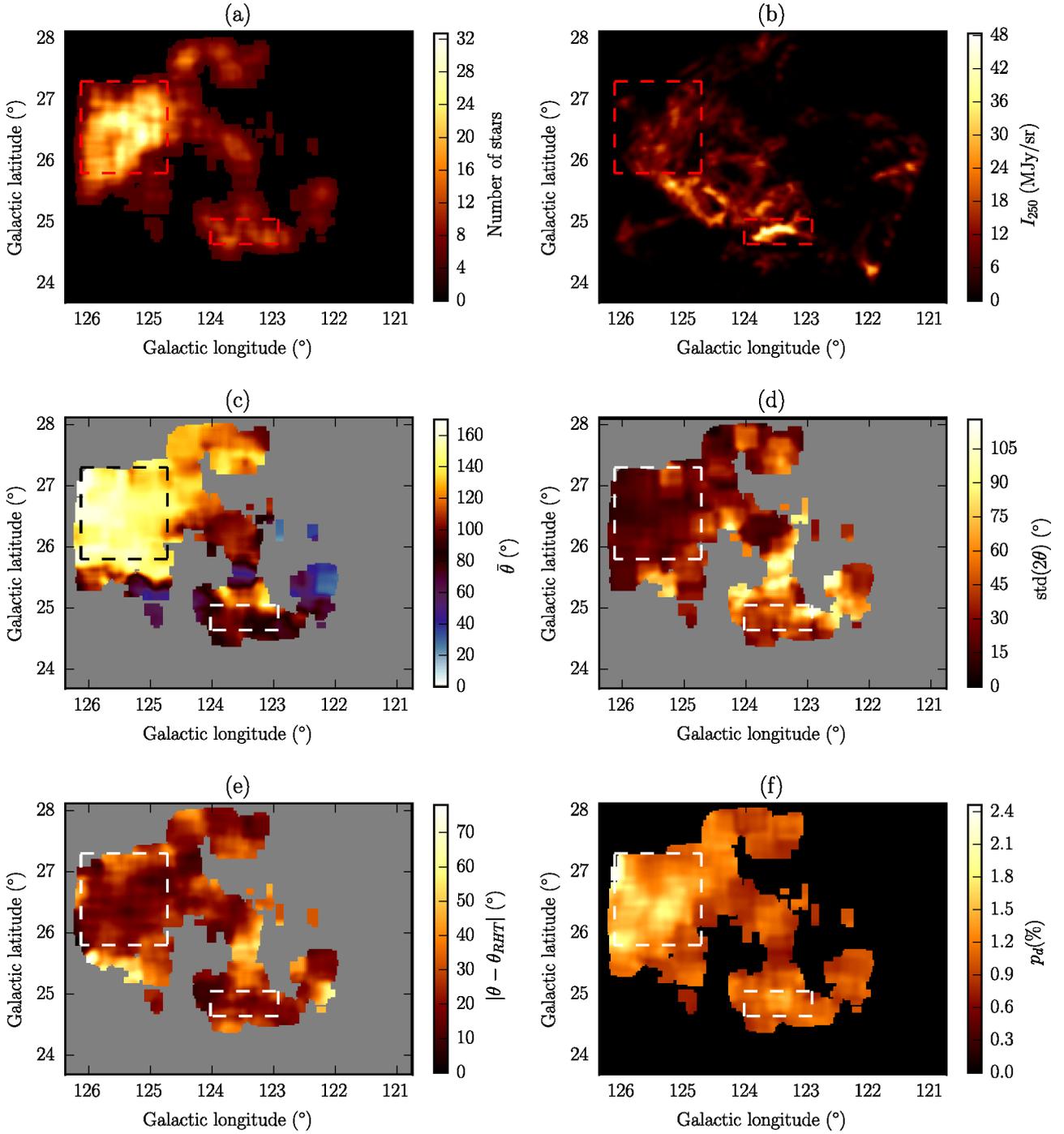}
\caption{Quantities smoothed on a 5\arcmin $\times$5\arcmin grid: (a) number of stellar polarization measurements, 
(b) \textit{Herschel} 250 $\rm\mu m$ intensity, (c) polarization angle, (d) polarization angle scatter, 
(e) relative orientation of $\rm B_{pos}$ and filaments, (f) fractional linear polarization. Rectangles outline regions A and B.}
\label{fig:smoothmaps}
\end{figure*}

The various trends existing in the measured cloud properties can be better visualized by constructing
maps of average (smoothed) quantities across the sky.
The maps are constructed by creating a grid (5\arcmin squares) of the field. The value at each grid center is 
calculated by averaging the values of star measurements within 10\arcmin of it. The final map is smoothed using a 
boxcar filter of 5\arcmin width. Fig. \ref{fig:smoothmaps} shows such maps of various quantities: (a) number of 
significant stellar polarization measurements ($p_d/\sigma_p \geq 2.5$), (b) \textit{Herschel} 250 $\rm\mu m$ 
intensity, (c) polarization angle, $\theta$, (d) scatter of $\theta$, (e) $\rm \vert\theta - \theta_{RHT}\vert$ and 
(f) $p_d$. Only bins with at least 5 measurements were used to produce the maps (reducing this number to 3 did not 
make a qualitative difference).

The number of significant stellar polarization measurements is highest in region A, which is the bright part
in the upper left of the map in panel (a). The number density (per 5\arcmin$\,$ bin) in this area is more than twice 
that in the rest of the map. Smaller local maxima are found: (i) in region B, (ii) in the center of the map
(123.5$^\circ$, +26.3$^\circ$) and (iii) above the striations at (124.5$^\circ$, +27.5$^\circ$). 
As noted in Paper I, this variation in the number of measurements is not due to variations in stellar density
across the field, or due to systematics such as observing conditions. 
Panel (b) shows the \textit{Herschel} 250 $\rm\mu m$ image smoothed with a boxcar filter of 5\arcmin width.
The MCLD123 filament (region B) stands out as a maximum of intensity.
By comparing the number density of significant stellar polarization measurements to the dust emission 
intensity (panels (a) and (b)) we find that the overall structure of the two maps is significantly different. 
Therefore, the variation of the density of significant polarization measurements cannot be attributed to 
variations in the column density.

The average polarization angles in panel (c) are calculated as the circular mean:
\begin{equation}
\centering
\bar{\theta} = \frac{1}{2} \, \rm arctan\left(\frac{\sum_{i = 1}^N \sin(2\theta_i)}{\sum_{i = 1}^N \cos(2\theta_i)}\right),
\end{equation}
where N is the number of measurements in each bin.
The average polarization angle defines a number of domains of different orientation throughout 
the field, separated by abrupt angle changes. 
The thin dark curves at the edges of some domains are an artefact of the smoothing, they do not represent
measurements of 90$^\circ$. $\bar{\theta}$ seems to rotate clockwise from 160$^\circ$ (bright yellow) in region A, 
to 40$^\circ$ (blue) below the center of the map.

The scatter of polarization angles $\rm std(2\theta)$ in panel (d) is measured using the circular standard
deviation:
\begin{equation}
\centering
\rm std(2\theta) = \sqrt{-2\ln\left(
\sqrt{\left(\frac{1}{N}\sum_{i = 1}^N \sin{2\theta_i}\right)^2 + \left(\frac{1}{N}\sum_{i = 1}^N \cos{2\theta_i}\right)^2}\right)}.
\label{eqn:std}
\end{equation}
The values obtained by this equation do not represent the dispersion that would characterize a normal distribution
of angles. They only serve for comparison between the various areas of the map.
The bright regions in panel (d) (large standard deviations) correspond to regions with very few measurements, 
as can be seen by comparing with panel (a). Region A and an area with $\rm std(2\theta) < 30^\circ$ in the center 
of the map stand out as the areas with lowest dispersion.

The relative orientation of $\rm B_{pos}$ and the filamentary structures in the cloud, shown in panel (e),
is very similar to the $\rm std(2\theta)$ map. Regions characterized by a large polarization angle dispersion 
appear to coincide with regions where filaments are perpendicular to $\rm B_{pos}$. 
The preference in alignment throughout the map, seen in Fig. \ref{fig:relativethetas}, is evident, as regions with 
$\rm \vert\theta - \theta_{RHT}\vert \lesssim 20^\circ$ (dark) occupy most of the map area.

\begin{figure*}
\centering
\includegraphics[scale=1]{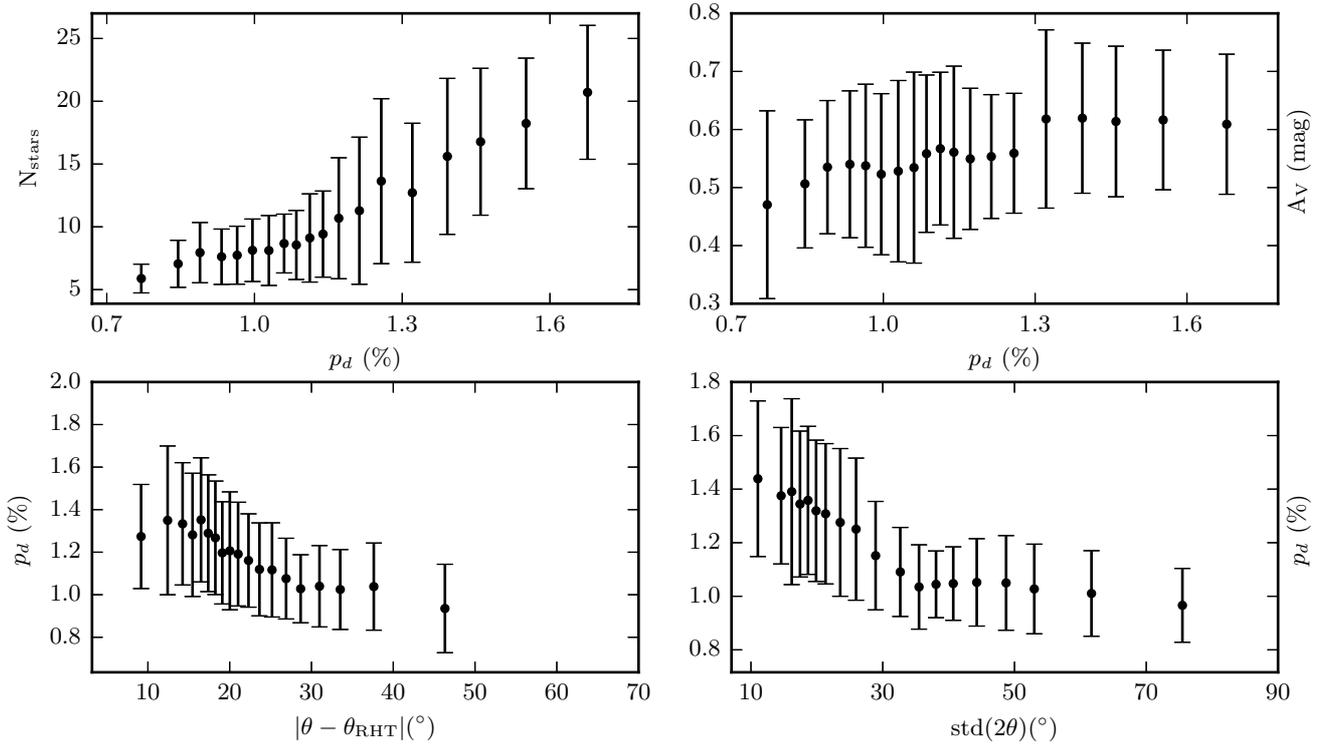}
\caption{Binned pixel-to-pixel comparison of maps from Fig. \ref{fig:smoothmaps}. 
Top left: $\rm N_{stars}$ versus $p_d$, top right: extinction $\rm A_V$ versus $p_d$, bottom left: 
$p_d$ versus $\rm \vert\theta - \theta_{RHT}\vert$, bottom right: $p_d$ versus $\rm std(\theta)$. 
Errorbars show the $\pm 1\sigma$ from the mean in each bin. Bins contain the same number of pixels.}
\label{fig:correlations}
\end{figure*}

The highest values of $p_d$ (panel f) are in the region of the striations. 
The \textit{Planck} polarized intensity peak coincides with the brightest part of this 
map at (126$^\circ$, +27$^\circ$) \citep[see figure A.1. in][]{planck20}. The second, but lower, local maximum in 
$p_d$ is in region B. The spatial variation of $p_d$ resembles that of $\rm N_{stars}$. This correlation is 
evident in Fig. \ref{fig:correlations} (top left), where the average value of $\rm N_{stars}$ is plotted for each 
(equally-populated with pixels) $p_d$ bin. This correlation is natural, since the polarization measurements were 
selected in Paper I to satisfy $p_d/\sigma_p \geq 2.5$. Since the observational errors $\sigma_p$ are uniform 
across the field (Paper I), it is more likely that a larger number of significant measurements will be obtained 
in regions of higher $p_d$.

Additional support to the fact that the variation in the number of significant polarization measurements is 
uncorrelated with the column density is given by the plot on the top right panel of Fig. \ref{fig:correlations}, 
where the visual extinction, $\rm A_V$, \citep[from the map of][]{cambresy} is plotted against $p_d$ (from panel f, Fig. \ref{fig:smoothmaps}). 
One possibility that explains the observed variation of $\rm N_{stars}$ 
is that the 3D orientation of the magnetic field varies significantly throughout the sky. If this is true, then it
is expected that more significant polarization will be detected in regions where the field is mostly
on the plane of the sky. In these regions, $p_d$ should be higher and this is consistent with the
correlation of $p_d$ and $\rm N_{stars}$ that we find. Also, if the field strength is more significant than
turbulent gas motions, which tend to randomize its orientation (see section \ref{ssec:Bstrength}), regions where 
$\rm B_{pos}$ is stronger (high $p_d$, $\rm N_{stars}$) should have a small angle scatter. This is indeed the case
as can be seen in the bottom right panel of Fig. \ref{fig:correlations}. However, a significant contribution to this trend is likely to be coming from the larger measurement error of lower-significance measurements (near the cut of $p_d/\sigma_p \geq 2.5$).

If the assumption of a strong magnetic 
field holds, then, diffuse, non self-gravitating filaments such as the ones in this cloud are 
expected to be mostly parallel to the 3D field orientation. So, if the field lies mostly on the plane of the sky, 
the alignment of the projected field and filaments will be easily detected. If, though, the field is mostly
along the line-of-sight, filaments and field can be observed as having any relative orientation. 
A hint of such a trend may exist in the bottom left panel of Fig. \ref{fig:correlations}.
\cite{planck32} find this exact trend (their figure 13) and attribute it to the same effect: the
projection of the 3D field on the sky. The anti-correlation of $p_d$ with $\rm std(\theta)$ is also observed  
\citep[figure 21 of][]{planck19}. 

\cite{planck32} also find that the degree of alignment decreases with column 
density in the range $\rm N_H$ = [$10^{20}, 10^{22}$] cm$^{-2}$. 
In contrast to this, it is evident even visually that panels (b) and (e) in Fig. \ref{fig:smoothmaps}
are uncorrelated $-$ we see no sign of such a trend. It is possible that the range of column densities in the Polaris
Flare is significantly different than that in the work of \cite{planck32}. To investigate whether this is the case,
we convert
the range of $\rm A_V$ in the Polaris Flare to $\rm N_H$ using the standard relation of \cite{bohlin}: 
$\rm N_H = 1.9 \times 10^{21} A_V \, (cm^{-2}/mag)$. From the $\rm A_V$ map of \cite{cambresy} values in the 
cloud vary within: $\rm A_V = [0.2, 1.2] \, mag \Rightarrow N_H = [0.4, 2.3] \times 10^{21} cm^{-2}$. 
Because the \cite{cambresy} data have a rather low resolution (8$\arcmin$), this range may be
underestimated. Indeed, the cores in MCLD123 have $\rm N_H \sim 10^{22} cm^{-2}$. 
Therefore, the range of $\rm N_H$ in the cloud is comparable to that of \cite{planck32}.
In their paper, the Planck collaboration infer that the anti-correlation of alignment with
column density is most likely due to the existence of molecular cloud structures that
are perpendicular to the magnetic field. The only available explanation for this 
observation is that matter in magnetically-dominated self-gravitating clouds collapses
preferentially along field lines, producing structures that are elongated and perpendicular
to the field. If we accept this reasoning, it should not come as a surprise that we do 
not find such a trend in the gravitationally unbound Polaris Flare.

In summary, the above correlations seem to indicate the presence of a strong magnetic field in the cloud which may change orientation from being mostly parallel to the plane-of-the-sky in region A
to being more inclined in other areas.

\subsection{Filament widths}
\label{ssec:width}
An important morphological characteristic of filaments is their width. 
We use the Filament Trait-Evaluated Reconstruction (FilTER) method\footnote{The code is available at: 
https://bitbucket.org/ginpan/filter} \citep{panopoulou2014} to construct the width distribution of filaments in the 
\textit{Herschel} image. 
FilTER takes as input the skeleton of an image and finds the width of the elongated structures. For this purpose,
a Gaussian is fit to the radial profile at every point along the filament axis, and the resulting FWHM is
deconvolved by the beam size. The skeleton is produced by DisPerSe \citep{sousbie} which is a 
topological code that can extract the filamentary structures of an image. 
It is designed to connect local maxima (cores). 
This property renders the construction of a representative skeleton difficult, because a significant part 
of the Polaris Flare does not contain bright cores. As seen in Fig. \ref{fig:rhtmap}, the RHT 
produces a visualization of filamentary structures irrespective of brightness. This enables us to apply DisPerSe to 
the RHT backprojection image and obtain a representative skeleton of the \textit{Herschel} image to use as input to
FilTER.

\begin{figure}
\centering
\includegraphics[scale=1]{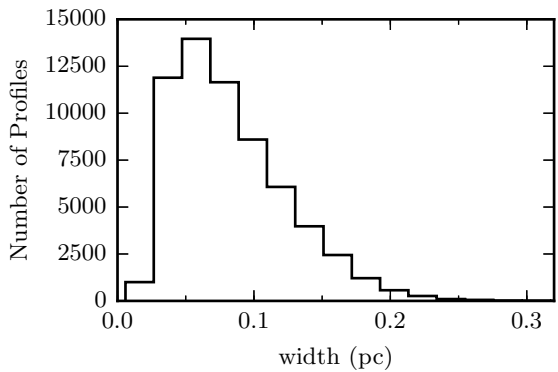}
\caption{Distribution of widths along filaments in the \textit{Herschel} image.}
\label{fig:widths}
\end{figure}

The distribution of filament widths in the \textit{Herschel} image is shown in Fig. \ref{fig:widths}. 
The distribution has a peak at 0.06 pc and a spread of $\rm \sigma_w = 0.04$, twice as 
much as the typical error in the width determination ($\rm \sigma_{fit}$) in our implementation of the code. 
The intrinsic spread of the distribution is: 
$\rm \sigma_{int} = (\sigma_w^2 - \sigma_{fit}^2)^{1/2} = 0.035 \,pc$.
Though lower than the characteristic width of 0.1 pc \citep[e.g.][]{arzoumanian2011,koch2015}, the peak value 
found here falls within the spread of the distributions from these works.

\begin{figure}
\centering
\includegraphics[scale=1]{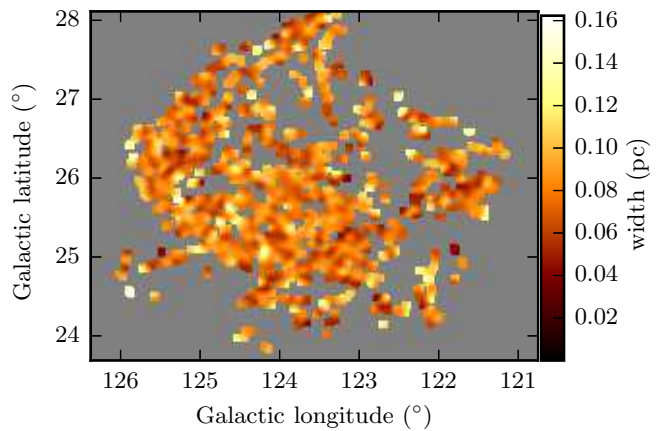}
\caption{Map of filament widths found with FilTER, smoothed with a 5\arcmin boxcar filter. }
\label{fig:widthmap}
\end{figure}

\begin{figure}
\centering
\includegraphics[scale=1]{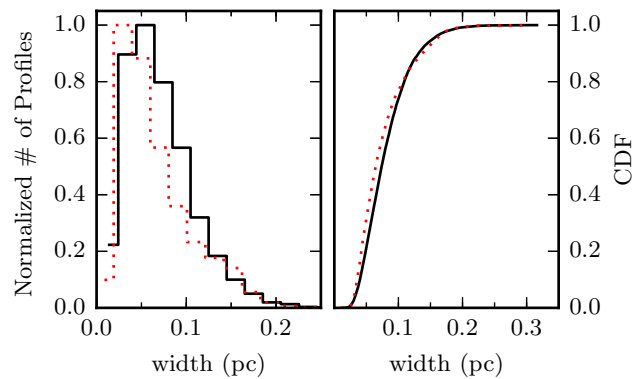}
\caption{Left: Normalized distribution of widths found with FilTER in the entire map (solid black) and in region 
A (dotted red) for all structures longer than 0.2 pc.  Right: CDF of the same distributions.}
\label{fig:widthsstriations}
\end{figure}

We investigate whether there are trends in the distribution of widths across the cloud, by constructing a smoothed
map of the widths found by FilTER (Fig. \ref{fig:widthmap}). We use a boxcar filter of the same size as the maps 
in Fig. \ref{fig:smoothmaps} (5\arcmin, or 0.2 pc) to allow for a better visual comparison.
The map shows fluctuations in the width of filaments that do not appear to have any large-scale systematic trend. 
This remains the case for maps with smaller smoothing kernels.
However, when comparing the distribution of widths within region A (Fig. \ref{fig:widthsstriations}, left, 
dotted red line) to that of the rest of the map (solid black line), we find that the former is slightly shifted
towards lower values. A KS test rejects the hypothesis that the two originate from the same parent distribution
at the 0.001 level. In constructing the distributions we discard structures with lengths less than 0.2 pc, as
these have too small aspect ratios to be considered filamentary. There is an indication that the shift towards 
lower widths becomes more pronounced as we raise this threshold to 0.4 pc.

\subsection{Analysis of regions A (striations) and B (MCLD123)}
\label{ssec:regions}
Having explored the various observed properties of the cloud and their variation across the map, we now focus
on the two regions of interest defined in Fig. \ref{fig:regions}.

\subsubsection{Region A (striations)}
\label{sssec:striations}
\begin{figure}
\includegraphics[scale=1]{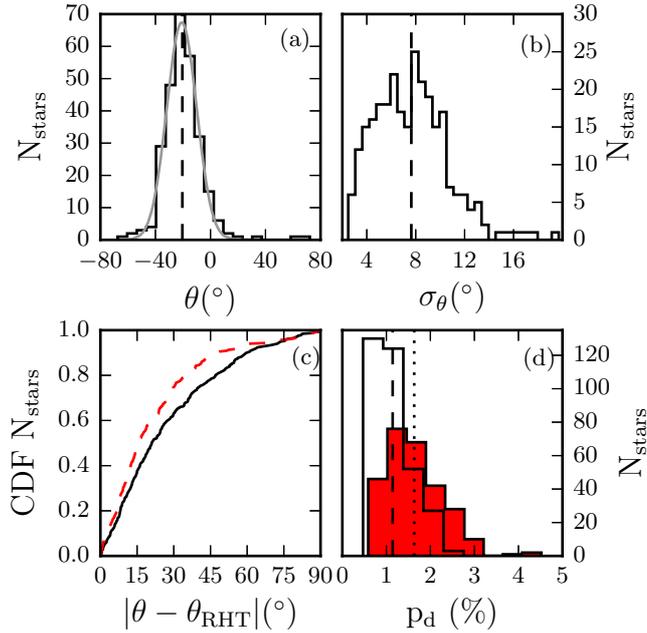}
\caption{(a): Polarization angle distribution in region A (black solid), Gaussian fit (gray), average (dashed). 
(b)  Distribution of polarization angle measurement errors in region A (solid) and mean (dashed). 
(c) Normalized CDFs of relative orientations of polarization measurements and dust striations in region A 
(dashed red) and the rest of the map (solid black). (d) Distributions of $p_d$ in region A (filled red) and the
rest of the map (empty black), dashed and dotted lines are their average values.}
\label{fig:striationplots}
\end{figure}

Region A contains striking features both in dust emission \citep[appendix of][]{miville} and in polarization
(Fig. \ref{fig:regions}). Both $\rm B_{pos}$ and striation orientations throughout the area are ordered. 
This is apparent in the distribution of polarization angles shown in panel (a) of Fig. \ref{fig:striationplots}.
The distribution of $\theta$ resembles a normal distribution with a mean at $-20^\circ$ and a standard deviation of 
$\rm \delta\theta_{obs} = 11^\circ$. The mean observational error (panel b) is 7.6$^\circ$ while the 
80$\rm^{th}$ percentile of the distribution is 10$^\circ$. Therefore, a significant contribution to the observed
spread is due to the measurement error.

In addition to this uniformity, the mean directions of the field and dust structures appear to be 
aligned and this occurs to a greater extent than in the rest of the map. This is supported by the comparison of the 
normalized CDFs of $\rm \vert\theta - \theta_{RHT}\vert$ shown in Fig. \ref{fig:striationplots}, panel (c). 
Inside the region (dashed red line) 75\% of the differences $\rm \vert\theta - \theta_{RHT}\vert$ lie in the range 
[0$^\circ$, 30$^\circ$], whereas outside (solid black) only 63\% are in the same range. 

Another characteristic of region A is that it exhibits a higher fractional linear polarization than the rest of the
map, as seen in panel (f) of Fig. \ref{fig:smoothmaps}. We compare the distribution of $p_d$ in region A (filled
red) and in the rest of the map (black empty) in panel (d) of Fig. \ref{fig:striationplots}. 
It is clear that $p_d$ in the region extends to higher values than out of the area, and that the mean $p_d$ 
(dashed line: out, dotted line: in) is lower outside the region.
 
\subsubsection{Region B (MCLD123)}

\begin{figure}
\includegraphics[scale=1]{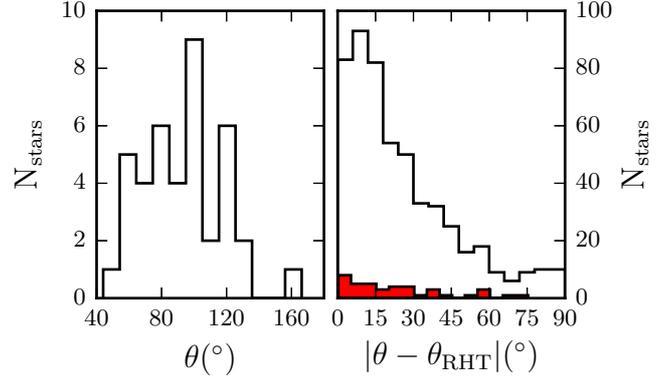}
\caption{Left: distribution of polarization angles in region B. Right: distribution of $\rm \vert\theta - \theta_{RHT}\vert$ for region B (filled red) and the rest of the map (empty black).}
\label{fig:MCLD}
\end{figure}

Region B shows a rather different picture than region A. In this area, which is the densest part of the cloud, 
the plane-of-the-sky field seems to bend along the MCLD123 filament.
The left panel of Fig. \ref{fig:MCLD} shows the distribution of $\theta$ for this region while the right 
panel shows the distribution of relative orientations, $\rm \vert\theta - \theta_{RHT}\vert$, for region B 
(filled red) and the rest of the map (empty black). The mean of the distribution of 
$\theta$ is 90$^\circ$, and the standard deviation is 25$^\circ$. The spread is partly due to the large-scale curvature of $\rm B_{pos}$ in the 
vicinity and on the MCLD123 filament. The values of $\rm \vert\theta - \theta_{RHT}\vert$ show a slight preference 
for alignment, although they extend to angles consistent with orthogonality. 

\subsection{Magnetic field strength and comparison to Turbulence}
\label{ssec:Bstrength}
The role of the magnetic field in the two regions discussed above can be investigated further by
estimating the strength of the plane-of-the-sky magnetic field and by comparing the magnetic energy
to that of turbulent motions. 

Polarization measurements can be used to estimate the strength of $\rm B_{pos}$ under the assumption that the 
polarization angle dispersion ($\delta \theta$) is caused by the action of hydromagnetic waves perpendicular to 
the (mean) magnetic field \citep{davis1951,chandrasekhar}(DCF). Based on the idea that a strong field will 
resist distortion and therefore, $\delta \theta$ will be small, DCF derive:
\begin{equation}
{\rm B_{pos}} \approx \sqrt{4\pi\rho}\frac{\delta v}{\delta \theta},
\label{eqn:DCF}
\end{equation}
where $\rho$ is the volume mass density of the gas and $\delta v$ is the velocity dispersion perpendicular to the 
observed field (along the line-of-sight). The last quantity, $\delta \theta$, can be expressed as the ratio of 
the (root-mean-squared) turbulent component of the magnetic field 
($\rm \left\langle B^2_t\right\rangle^{1/2}$) to the ordered component ($\rm B_0$)\citep{hildebrand}:
\begin{equation}
\delta \theta \approx \rm \frac{\left\langle B^2_t\right\rangle^{1/2}}{B_0}.
\label{eqn:dthetaratio}
\end{equation}

\subsubsection{The turbulent-to-ordered field ratio}
\label{sssec:BtBo}
Following \cite{hildebrand} an estimate of the relative strength of $\rm B_0$ with respect to 
$\rm \left\langle B^2_t\right\rangle^{1/2}$ can be obtained by calculating the dispersion function of the
polarization angles, defined as:
\begin{equation}
\left\langle \Delta \theta^2(l) \right\rangle_{tot} = \frac{1}{N(l)} \sum_{i=1}^{N(l)}{\Delta\theta_i^2(l)}
\label{eqn:DF}
\end{equation}
where the angle differences of all (unique) pairs of polarization measurements, $ \Delta \theta_i(l)$, 
are binned according to the angular distance $l$ in bins containing $N(l)$ pairs. The angle difference of 
the $i^{th}$ pair, is simply the difference between the polarization angle measured at the position 
$x_i$ and that at $x_i+l$: $\Delta \theta_i(l) = \theta(x_i) - \theta(x_i+l)$. 
We constrain $ \Delta \theta_i(l) \in [0^\circ,90^\circ]$.  
The dispersion function, corrected for the scatter induced by measurement errors ($\sigma_M$), is expected to 
follow the equation:
\begin{equation}
\left\langle \Delta \theta^2(l) \right\rangle_{tot} - \sigma^2_M \simeq m^2l^2 + b^2(1-e^{-l^2/2\delta^2}),
\label{eqn:dthetatot}
\end{equation}
as shown by \cite{houde2009} \citep[their equation 44 adapted for optical polarimetry data by][]{franco2010}. 
In equation \ref{eqn:dthetatot}, $\delta$ is the correlation length of $\rm B_t$, $b$ and $m$ are constants of 
proportionality, and $\sigma^2_M(l)$ is found according to the equation:
\begin{equation}
\sigma^2_M(l) = \frac{1}{N(l)} \sum_{i=1}^{N(l)} \sigma^2_{\Delta\theta_i(l)},
\label{eqn:sigmam}
\end{equation}
where $\sigma_{\Delta\theta_i(l)}$ results from error propagation. Equation \ref{eqn:dthetatot} is expected 
to hold for distances smaller than the typical scale ($d$) for large scale variations in $\rm B_0$. 
\begin{figure*}
\centering
\includegraphics[scale=1]{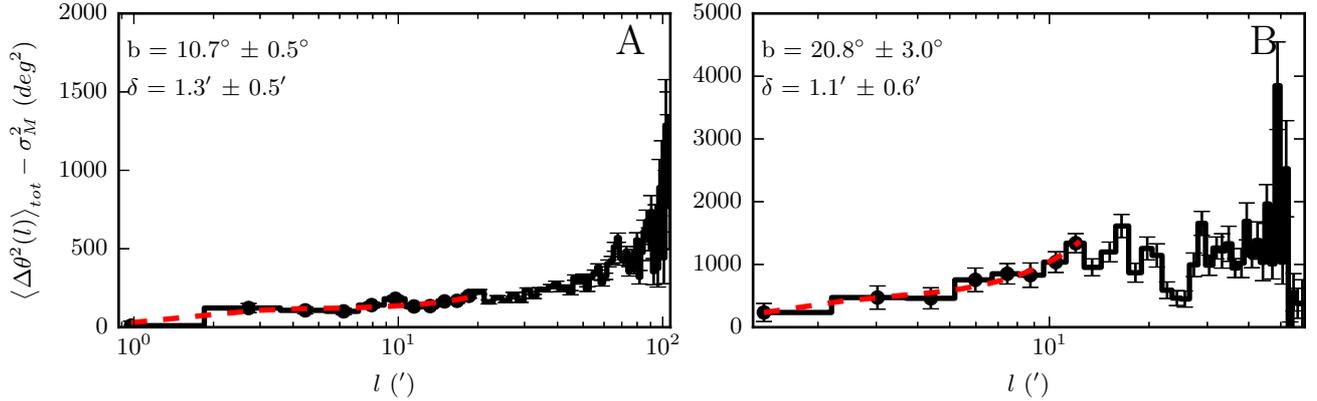}
\caption{The dispersion function (corrected for measurement errors) vs. angular distance, $l$, 
for region A (left) and B (right).
The red dashed line shows the best fit function of the form 
$\left\langle \Delta \theta^2(l) \right\rangle_{tot} -\sigma_M^2 = m^2l^2+ b^2(1-e^{-l^2/2\delta^2})$. 
The black dots show the data points that were used in the fits. The fit results for $b$ and $\delta$ 
are shown in the top left.}
\label{fig:SFplots}
\end{figure*}
The parameter $b$ is related to the ratio of the rms $\rm B_t$ to $\rm B_0$ by:
\begin{equation}
b^2 = \frac{2}{\mathcal{N}} \rm \frac{\left\langle B^2_t\right\rangle}{B^2_0},
\label{eqn:bhoude}
\end{equation} 
where $\mathcal{N}$ is the number of turbulent cells along the line-of-sight:
\begin{equation}
\mathcal{N} \simeq \rm \frac{s_{los}}{\sqrt{2\pi}\delta},
\label{eqn:N}
\end{equation} and $\rm s_{los}$ is the line-of-sight dimension of the cloud.

We implement the method of \cite{hildebrand} and \cite{houde2009}(H09) in regions A and B. 
We calculate the dispersion function 
$\left\langle \Delta \theta^2(l) \right\rangle_{tot}$, as explained above, correct it to obtain the left side of 
equation \ref{eqn:dthetatot} and fit the function on the right side of equation \ref{eqn:dthetatot} to these 
values with respect to $l^2$. 
The fit takes into account the error of $\left\langle \Delta \theta^2(l) \right\rangle_{tot}$. 
In Fig. \ref{fig:SFplots} we plot $\left\langle \Delta \theta^2(l) \right\rangle_{tot}- \sigma^2_M$ with distance, 
$l$, for each region (black solid lines). The black dots in Figure \ref{fig:SFplots} show the values used for the 
fit. The fits $p(l) = m^2l^2 + b^2(1-e^{-l^2/2\delta^2})$ are shown with red dashed lines. 
The distances are shown on a logarithmic scale (the bins used for fitting were linear). 
The bin size in region B (1.5\arcmin) was chosen so that there were more than 10 points per bin 
(in the fitting range). The bin size (1.7\arcmin) for region A was the minimum that produced a positive value of 
$\left\langle \Delta \theta^2(l) \right\rangle_{tot} - \sigma^2_M$ for the first bin. 
A choice of bin size larger by up to 38\% in region A and 33\% in region B, produces values of
$b$ and $\delta$ within the error of the fit.
The fits were done considering only distances smaller than the scale at which $\rm 
B_{pos}$ varies ($l < d$). In region A we performed the fits up to $l \approx 20'$. In region B we fit the data up 
to $l \approx 13'$. 
The choice of $d$ is that which permits the maximum number of bins to be used in the fit while 
providing a fit that traces the exponential cut-off as well as possible. In region A both $b$ and $\delta$ remain 
within the error of the fit for a choice of $d$ in the range $8\arcmin-25\arcmin$. The same holds for region B
in the range $9\arcmin - 13\arcmin$.

The results of the fits are shown in the top left of the panels in Fig. \ref{fig:SFplots}. The correlation length of $\rm B_t$ is not well constrained, as only in the first bin of region A the effect of the
exponential decay in equation \ref{eqn:bhoude} is apparent. The values for $\delta$ obtained for both regions
are consistent within the errors of the fits. Inserting the value for $b$ and $\delta$ from the fits into equation 
\ref{eqn:bhoude} we obtain estimates for the ratio $\rm \left\langle B^2_t\right\rangle^{1/2}/B_0$:
0.2$-$0.5 for region A and 0.3$-0.8$ for region B.
The line-of-sight depth of each region used for these estimates is presented and discussed in the next section.

Within our adopted uncertainties, the ratios in both regions are less than 1. This indicates that the 
magnetic field is much stronger than the turbulent component in region A and at most comparable to the turbulent 
component in region B.
Moreover, equation \ref{eqn:DCF} can be written as:
\begin{equation}
{\rm \frac{\left\langle B^2_t\right\rangle^{1/2}}{B_0}} \simeq \frac{\delta v}{v_A}
\label{eqn:beta}
\end{equation}
\citep{hildebrand}, where $v_A = \rm B_0/\sqrt{4\pi \rho}$ is the Alfv\'en speed and $\delta v$ the velocity 
dispersion perpendicular to the mean field. From this we can conclude that turbulent motions 
are most likely sub-Alfv\'enic in region A and could be sub- to trans-Alfv\'enic in region B.
This is consistent with the conclusion of \cite{hily-blant2007} that the magnetic field is dynamically important 
in region B. These authors found that motions in MCLD123 are trans-Alfv\'enic, using the dispersion of position 
angles of diffuse $\rm ^{12}CO$ filaments.

\subsubsection{The strength of the ordered component}
\label{sssec:Bstrength}

Having inferred the value of $\delta\theta$ from the H09 method, we now move on to find the values of the remaining
quantities that enter into equation \ref{eqn:DCF} and use them to obtain an estimate for $\rm B_{pos}$ in both 
regions. We present the values used in obtaining these estimates in Table \ref{tab:DCF}.

The gas mass density is equal to: $\rho = \rm \mu m_H n_H$, where $\rm m_H$ is the mass of the hydrogen atom, 
$\mu = 1.36$ is a factor that accounts for the fraction of helium and $\rm n_H$ is the total hydrogen number 
density. The total $\rm n_H$ must be used and not $\rm n_{H_2}$ because even though the cloud is molecular, the 
fraction of gas that is in atomic form may be important \citep[$\rm N_{HI} \sim N_{H_2}$][]{hily-blant2007}. 
To estimate the density in region B, we assume that MCLD123 has a cylindrical geometry. This implies
a line-of-sight dimension of $\sim 0.25- 0.5$ pc, where the lower bound is twice the observed extent of the dense
part of the filament and the upper bound is found including the more diffuse parts. 
The average $\rm N_{H_2}$ from the map of \cite{andre2010} is $\rm 8 \times 10^{20} \,cm^{-2}$, 
corresponding to an $\rm N_H \sim 2.4 \times 10^{21} \,cm^{-2}$. 
Dividing by the assumed line-of-sight dimension gives $\rm n_H \simeq 1500 - 3100 \,cm^{-3}$ \citep[see also][]{grossman}.

We can obtain a loose constraint on the mean density of region A by arguing that it is 
likely to be less than that of region B (since the latter contains cores). An upper bound would therefore be 
$\rm n_H^A < n_H^B$. A lower limit on the density can be estimated by
considering existing measurements of $\rm n_H$ from UV data of stars in diffuse sightlines. A set of measurements 
from various works has been collected by \cite{goldsmith2013} who shows that most sightlines with 
$\rm N_H \gtrsim 10^{21} \, cm^{-2} \Rightarrow N_{H_2} \gtrsim 3 \times 10^{20} \, cm^{-2} $ have measured 
densities: $\rm 300 \,cm^{-3} \gtrsim n_H \gtrsim 60\, cm^{-3}$ (assuming $\rm N_{HI} \simeq N_{H_2}$).
The column density in region A is estimated\footnote{this is consistent 
with the value obtained by converting the mean $\rm A_V$ to $\rm N_H$ according to \cite{bohlin}: 
$\rm N_H = 1.9 \times 10^{21} A_V \, (cm^{-2}/mag)$ for $\rm R_V = 3.1$ \citep[e.g.][]{savagereview}} by 
$\rm N_H^A \simeq N_H^B \, I_{250}^A/I_{250}^B \simeq 1.2 \times 10^{21} cm^{-2}$, 
where we have assumed that the temperature distribution is quasi-uniform throughout the cloud 
\citep{menshchikov,schneider2013} so that intensity ($\rm I_{250}$) variations in the \textit{Herschel} image are 
mostly caused by $\rm N_H$ fluctuations. Since the Polaris Flare is a molecular cloud, we reason that $\rm n_H^A > 
300 \,cm^{-3}$ (the upper limit from the aforementioned sightlines) is an appropriate 
approximation. The line-of-sight dimension of region A is then ($\rm s_{los} = n^A_H/N^A_H$) $\rm 1.3 \, pc \gtrsim s_{los} \gtrsim 0.2 \, pc$. 
The upper limit is approximately one third of the size of the region as projected on the sky, 
while the lower limit is similar to the assumed $\rm s_{los}$ in region B. 
We expect that the true $\rm s_{los}$ lies mostly near the lower limit
of this range, because due to the translucent nature of the cloud, the line-of-sight dimension should not 
be very large. Studies of cirrus clouds in general find a high likelihood of them being 
sheet-like \citep[e.g.][]{gillmon2006}. Along with the fact that regions A and B have similar mean $\rm A_V$ 
(Table \ref{tab:DCF}), this implies that $\rm s_{los}$ in the two regions should not vary by orders of magnitude.
 
\begin{table*}
\centering
\caption{Values used for estimation of $\rm B_{pos}$ in regions A and B. 
$\Delta V$: FWHM of CO (J=1-0) line, 
$\rm A_V$: average visual extinction \citep{cambresy}, 
$\rm N_H$: hydrogen column density, 
$\rm s_{los}$: line-of-sight dimension, 
$\rm n_H$: hydrogen number density, 
$\delta \theta^{H09}$: polarization angle dispersion from H09 method (equation \ref{eqn:dthetaratio}), 
$\delta \theta$: polarization angle dispersion from section \ref{sssec:striations} 
(corrected for measurement error according to equation \ref{eqn:dthetaobs}),
$\rm \left \langle B^2_t \right \rangle^{1/2}/B_0$: turbulent-to-ordered field ratio from H09 method,
$\rm B^{H09}_{pos}$: plane of sky magnetic field using $\rm \delta \theta^{H09}$ and equation \ref{eqn:DCF},
$\rm B^{mDCF}_{pos}$: plane of sky magnetic field using $\delta \theta$
and the modified equation \ref{eqn:DCF} as explained in the text.}
\begin{tabular}{|c|c|c|c|c|c|c|c|c|c|c|c|}
\hline
Region & $\Delta V$ & $\rm A_V$ & $\rm N_H$ & $\rm s_{los}$&$\rm n_H$  & $\rm \delta \theta^{H09}$ &$\delta \theta$ &$\frac{\sqrt{\left\langle \rm B^2_t\right\rangle}}{B_0}$ &$\rm B^{H09}_{pos}$&$\rm B^{mDCF}_{pos}$\\
       & (km/s)     & (mag) & $\rm 10^{20}(cm^{-2})$ & (pc) &$\rm (cm^{-3})$& $(^\circ)$ &$(^\circ)$ & &($\mu$G)&($\mu$G)\\
\hline
\hline
  A    & 3.1 & 0.6 & 12 & $0.4 - 1.3$ & $300-1000$ & $10^\circ-29^\circ$ & 8 &$0.2-0.5$& $24-120$& $43-81$\\
\hline
  B    & 2.6 & 0.7 & 24 & $0.25-0.5$ &$1500-3100$& $17^\circ-44^\circ$ & & $0.3-0.8$ & $30-111$&\\
\hline
%\footnotesize{$^*$ $\rm s_{los} = N_H/n_H$.}
\end{tabular}
\label{tab:DCF}
\end{table*}
%Another estimate can be obtained by the dust-inferred column density shown in figure 1 of \cite{andre2010}.
%Their $N(H_2)$ map covers the central part of the Polaris Flare field and does not contain region A. 
%However, assuming that variations in dust emission intensity are mainly due to variations in column density, 
%the $N(H_2)$ in the striations is roughly equal to $N(H_2)$ of a region with similar intensity in the 
%$250 \, \mu m$ image.

To estimate the velocity dispersion, $\delta v$, we use the $\rm ^{12}CO$ (J = $1-0$) data from the survey of 
\cite{dame2001}\footnote{Survey online archive: \\https://www.cfa.harvard.edu/rtdc/CO/IndividualSurveys/}. 
The data, first presented by \cite{heithausen1993}, cover 134 deg$^2$ %Dame et al table 1
including the \textit{Herschel}-mapped area. The angular resolution is 8.7\arcmin and the spectral
resolution is 0.65 km s$^{-1}$. We fit a Gaussian to the mean spectrum of each region and relate its 
FWHM ($\Delta V$) to the velocity dispersion ($\delta v$) with: $\delta v = \Delta V/(2 \sqrt{2\ln2})$.
Complications may arise when the $\rm ^{12}CO$ (J = $1-0$) is used to estimate the velocity dispersion 
if the line is optically thick. However, \cite{hily-blant2007} found that the $\rm ^{12}CO$ (J = $1-0$) line is optically 
thin in diffuse gas within the MCLD123 filament with column density derived from the scaling of CO to $\rm ^{13}CO$ 
velocity-integrated line temperatures: $\rm N_{H_2} \sim 10^{20} cm^{-2}$. Since the mean column density in region A
is $\rm N_{H_2} \approx 4 \cdot 10^{20} cm^{-2}$ ($\rm N_{HI} \approx N_{H_2}$) this indicates that the line 
may also be optically thin within region A. The line is most likely optically thick in the densest parts of 
region B, however these occupy a very small fraction of the area considered.

The results that follow from equation \ref{eqn:DCF} are presented in Table \ref{tab:DCF}.
We show the entire range of values, given the possible variation in the estimated quantities. The range arises 
from two factors: the error on $\delta$ and the estimate of $\rm s_{los}$, which enters both in the 
estimation of the density $\rho$ and the dispersion of polarization angles from the H09 method.
The $\rm B_{pos}$ estimates between regions A and B are very similar.

\cite{hily-blant2007} obtained an estimate of $\rm B \simeq 15 \,\mu G$ for a smaller part of region B, 
using the angle dispersion of filaments seen in $\rm ^{12}CO$. This value is 50\% that of the 
lowest bound of our estimate. 

\subsubsection*{$\rm B_{pos}$ with alternative estimate of $\delta \theta$.}
We can obtain an alternative estimate of $\rm B_{pos}$ by using the observed angular 
dispersion $\rm \delta \theta_{obs}$ found in section \ref{sssec:striations} from the Gaussian fit to the 
distribution of polarization angles \citep[see for example section 4.2.3 of][] {barnes}.
\cite{ostriker} applied the DCF method to numerical simulations of MHD turbulence using this type of 
angle dispersion calculation and found that the method provides a good estimate of $\rm B_{pos}$ when 
$\delta \theta \lesssim 25^\circ$. They introduced a factor $f$ in equation
\ref{eqn:DCF} to correct for line-of-sight averaging: ${\rm B_{pos}} = f \sqrt{4\pi\rho} \,\delta v/\delta \theta$
and proposed $f \approx 0.5$.

We use this modification of equation \ref{eqn:DCF} with $\rm \delta \theta_{obs}$ to obtain another 
estimate of $\rm B_{pos}$ in region A. 
Because the errors in polarization angle ($\sigma_\theta$) add to the intrinsic angle dispersion of the cloud, 
$\delta\theta$, we correct for this bias by \citep[e.g.][]{crutcher2004,girart}:
\begin{equation}
\delta \theta^2 = \delta \theta_{obs}^2 - \bar{\sigma}_\theta^2,
\label{eqn:dthetaobs}
\end{equation} 
where $\bar{\sigma}_\theta$ is the mean polarization angle error.

The values for the magnetic field strength are shown in the last column of Table \ref{tab:DCF}.
They lie within the range of values found using the angle dispersion from H09 and the original equation \ref{eqn:DCF}.

\section{Discussion}
\label{sec:discussion}

\subsection{Properties of the 3D magnetic field}
The projected magnetic field of the cloud presents a very inhomogeneous structure. There are regions where
it is uniform and others where the measured orientations appear random, or significant measurements are 
entirely absent. These characteristics may provide hints on the nature of the 3D field.
Let us consider region A, which has the largest density of significant polarization measurements. 
As discussed in Paper I, this is not due to variation in stellar density, observing conditions or other 
systematics. Therefore, region A must be characterized by higher polarization efficiency (in the terminology of
\cite{andersson2015}: intrinsic polarization per unit column density). 
We can further investigate this observation by considering that $p_d$ is related to the following 
factors \citep{leedraine}: 
\begin{equation}
p_d = p_0 R F \cos^2\gamma . 
\label{eqn:leedraine}
\end{equation}
where $\gamma$ is the inclination angle (the angle between the magnetic field and the plane-of-the-sky), $p_0$ 
reflects the polarizing capability of the dust grains due to their geometric and chemical characteristics, 
and $R$ is the Reyleigh reduction factor \citep{greenberg} which quantifies the degree of alignment of the grains 
with the magnetic field. $F$ accounts for the variation of the field orientation along the line-of-sight and is
equal to $F = \frac{3}{2}(\left\langle \cos^2\chi\right\rangle-\frac{1}{3})$, where $\chi$ is the angle between 
the direction of the field at any point along the line-of-sight and the mean field direction. The angular brackets 
denote an average along the line-of-sight.
The increased $p_d$ of region A could therefore be the result of any of these factors (or some combination
of them), in other words region A could have:
\begin{itemize}
\item[i)] increased alignment efficiency (e.g. more background radiation, a larger amount of asymmetric dust grains, larger grain sizes), 
i.e. higher factors $p_0$ and/or $R$,
\item[ii)] more uniform magnetic field along the line-of-sight, i.e. higher $F$,
\item[iii)] increased $\rm B_{pos}$ (inclination of B is larger with respect to the line-of-sight), i.e. higher $\cos\gamma$.
\end{itemize}
Several pieces of evidence challenge the validity of case (i). First, if the radiation illuminating region A
were much different in intensity or direction, then the dust temperature of the area would have to be 
qualitatively different (higher) than in other regions of the cloud. As mentioned in section \ref{ssec:Bstrength},
the results from \cite{menshchikov} imply that temperature and density variations are
subtle across the field. Indeed, the temperature PDF presented by \cite{schneider2013} is narrow. 
Also, the most likely candidate for providing illumination to the cloud, due to its likely proximity, 
is Polaris (the star). \cite{zagury1999} concluded by studying optical and $\rm 100\,\mu m$ 
light from MCLD123, that the star cannot be the primary source of dust heating. 
Additionally, larger amounts of (aligned) grains
would imply larger column densities (or $\rm A_V$) than the rest of the cloud, which is not a characteristic of 
region A. To the best of our knowledge, evidence for significant variation of grain size within the same cloud 
between regions of such similar $\rm A_V$ does not exist. 

We now investigate whether the observed difference in $p_d$ between region A and other parts of the 
cloud could arise from differences in the properties of the magnetic field along the line-of-sight.
We can obtain an upper limit on the influence of the factor $F$ on $p_d$ by keeping all other factors in equation
\ref{eqn:leedraine} constant and taking the ratio of two regions, for example A and B:
\begin{equation}
\frac{p_A}{p_B} \approx 
\frac{\left\langle \cos^2 \chi_A \right\rangle - \frac{1}{3}}{\left\langle \cos^2 \chi_B \right\rangle -\frac{1}{3}}
\label{eqn:pratioF}
\end{equation}
where the average $p_d$ in regions A and B are equal to $p_A = 1.63, \, p_B = 1.3$ and
$\left\langle \cos^2{\chi_i} \right\rangle$ 
($i = A, B$) is an average over all lines of sight in region $i$. For region A, the angle dispersion is small 
($\delta\theta \sim 10 ^\circ$, section \ref{ssec:regions}), so we can make the approximation $\left\langle 
\cos^2{\chi_A} \right\rangle \approx \cos^2{10^\circ}$ (to better than 10\%). We check what values of 
$\left\langle \cos^2{\chi_B} \right\rangle$ produce the observed ratio of $p_A/p_B$ (within 30\%) 
by drawing $\mathcal{N} = 1-9$ angles (the number of turbulent cells in region B) from a normal distribution with 
$\sigma = 10^\circ - 50^\circ$. 
After repeating the process 100 times, we find that the most likely $\sigma$ that can reproduce the observed
$p_A/p_B$ are in the range $10^\circ-25^\circ$, similar to the observed angle dispersion on the 
plane-of-the-sky (section \ref{ssec:Bstrength}). Therefore, it is possible that region A has a more ordered field along the 
line-of-sight compared to other regions.

Finally, if the 3D orientation of the magnetic field is mostly in the plane-of-the-sky in region A 
and less so in other parts of the cloud, this could also explain the increased measurement density in this region.
We can estimate the change in angle that is needed to obtain the difference in $p_d$ between regions A and B, 
from equation \ref{eqn:leedraine}.
Taking all factors equal between the two regions except the inclination angle, the ratio of $p_d$ is: 
$p_d^A/p_d^B = 1.63/1.3 = \cos^2\gamma_A/\cos^2\gamma_B$. This ratio could arise either from a pair of
large $\gamma_A,\, \gamma_B$ with a small difference or from small angles having a large difference. 
This ambiguity can be lifted by considering the expected polarization angle dispersions for different
inclinations of the magnetic field, studied with MHD simulations by \cite{falceta}. 
In their work, these authors found that in the range
$0^\circ-60^\circ$ the predicted polarization angle dispersions are below $35^\circ$ for their model with 
a strong magnetic field (Alfv\'en Mach number 0.7). 
Since the angle dispersions in the two regions are significantly lower than this value, it is safe to assume 
that $0^\circ < \gamma_A,\gamma_B < 60^\circ$. For this range, we find that the observed ratio of $p_d$ 
can be explained by inclination differences most likely in the range $\gamma_B-\gamma_A = 6^\circ - 30^\circ$.
In order to explain the difference in $p_d$ between region A and the lowest mean-$p_d$ region in the map 
(Fig 4, panel f, approximately at the center of the map) which has a $p_d \sim 0.7\%$, 
the difference in inclination angle needs to be $20^\circ-50^\circ$.

With the existing set of measurements, we are not able to conclude whether the variations
in polarization fraction are mostly due to change in inclination or due to differences in the uniformity of 
the field along the line-of-sight. We have, however, provided bounds on the areas of the parameter 
space in which these differences are likely to occur.

\subsection{Cloud distance}
\begin{table}
\centering
\caption{Distance estimates to the cloud from different references.}
\begin{tabular}{|c|c|c|}
\hline
Reference & $d$ (pc) & method \\
\hline
\hline
\cite{heithausen1990} & $\leq 240$   & Stellar extinction,\\ && nearby clouds\\
\cite{zagury1999}     & 125 $\pm$ 25 & Association to stars \\
\cite{brunt2003}      & 205 $\pm$ 62 & Size-linewidth relation \\
\cite{schlafly2014}  & 390 $\pm$ 34 & Stellar extinction  \\
\hline
\end{tabular}
\label{tab:distances}
\end{table}
In the literature there is no definitive consensus on the cloud's distance. The existing estimates are shown
in table \ref{tab:distances} along with the method that was used to obtain each one. 

\cite{heithausen1990} base their distance estimate on reddening estimates of stars in the field from 
\cite{keenan}, who found
reddened stars from distances as close as 100 pc and that all stars farther than 300 pc were reddened. 
Knowing that Polaris (the star) showed dust-induced polarization, and that the then existing distance estimates to
the star were 109 - 240 pc they placed the cloud at a distance $d < 240$ pc. This distance fits with the smooth
merging of the cloud at lower latitudes with the Cepheus Flare, at 250 pc.
\begin{figure}
\centering
\includegraphics[scale=1]{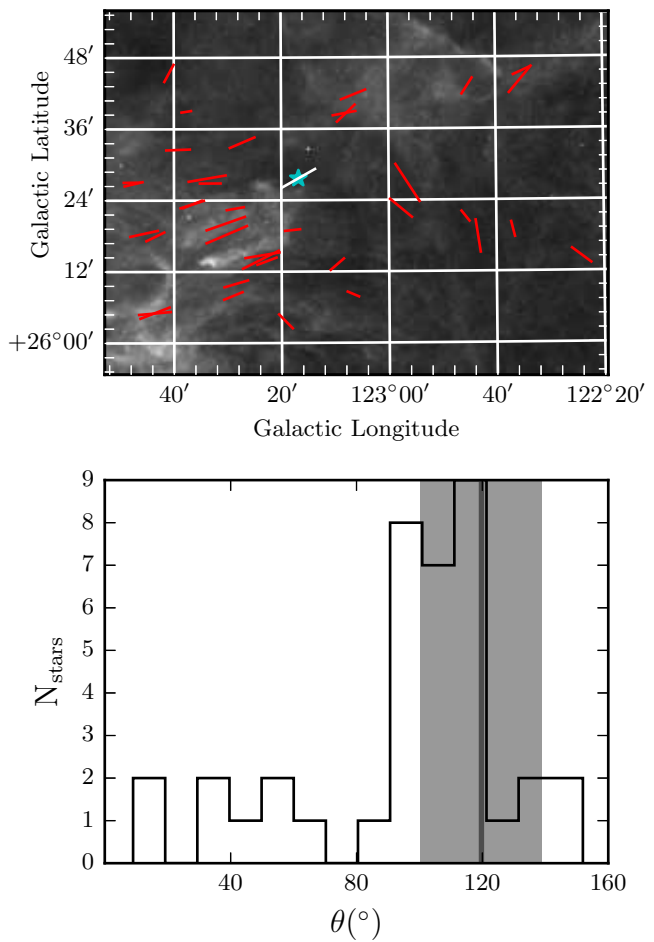}
\caption{Top: Polarization angle of the star Polaris \citep{heiles2000}, white segment, compared to our data,
 red segments. The $p_d$ of Polaris is 0.1\%, but has been enhanced 10 times. Its position is marked with a star.
 Bottom: Distribution of $\theta$ in the area shown in the top panel. The dark gray line shows the $\theta$ of 
 Polaris while the gray band is the 1$\sigma$ error. Bin size is 10$^\circ$.}
\label{fig:northstar}
\end{figure}

\cite{zagury1999} compared IRAS 100 $\rm\mu m$ emission with optical images of MCLD123 and found that the 
brightness ratios are consistent with Polaris (the star) being the illuminating source of the cloud in the 
optical. They placed the cloud at a distance 6 $-$ 25 pc in front of the star ($\rm 105 \, pc < d < 125\, pc$ from the sun) so that its contribution in dust heating would be minimal compared to the interstellar radiation field.

\cite{brunt2003} used Principal Component Analysis (PCA) of spectral imaging data to infer distances based on the 
universality of the size-linewidth relation for molecular clouds. Their estimate is consistent with both
the above estimates.

\cite{schlafly2014} used accurate photometry measurements of the Pan-STARRS1 survey and 
calculated distances to most known molecular clouds. Their estimate for the distance of the 
Polaris Flare (390 pc), was obtained for lines of sight in the outskirts of the cloud (outside our observed field). 

In Fig. \ref{fig:northstar} (top) we zoom in on the region surrounding the North Star and overplot the polarization
data (red) and the measurement of the North Star (white) from the \cite{heiles2000} catalogue on the 
\textit{Herschel} image. The lengths of the segments are proportional to their $p_d$, and the length of the Heiles 
measurement has been enhanced 10 times. The position of the North Star is marked with a blue star. 
The star happens to be projected on an area of very little dust emission, hence the low $p_d$. 
Magnetic field orientations in the area show a strong peak around $110^\circ$, with a few
measurements (that happen to fall towards the right and bottom of the area) clustering around $40^\circ$. 
This can be seen in the distribution
of angles in the area shown in Fig. \ref{fig:northstar} (bottom). The polarization angle of the North Star 
\citep{heiles2000} is shown with the dark gray vertical line and the light gray band shows the 1 $\sigma$ error.  
The stars that are nearest to Polaris, in projection, belong to the peak at 100$^\circ$.
The fact that the orientation of the North Star's polarization is consistent with this peak 
is intriguing. It could add to the evidence supporting that Polaris is behind the Flare, constraining 
its  distance to the \cite{zagury1999} estimate. However, since the orientation of stellar polarization 
shifts by a substantial amount (60$^\circ$) in $\sim 20'$, denser sampling of the area is needed  
in order to ascertain this indication.

\section{Summary}
\label{sec:summary}
We combined RoboPol optical polarization measurements and \textit{Herschel} dust emission data to infer the 
magnetic field properties of the Polaris Flare. We found that linear dust structures (filaments and striations) 
are preferentially aligned with the projected magnetic field. This alignment is more prominent in regions where 
the fractional linear polarization is highest (and the number of significant polarization measurements is 
largest). This correlation supports the idea that variations in the alignment are partly caused by the 
projection of the 3-dimensional magnetic field. We investigated the possibility of important spatial variations 
in the filament widths and found only a slight indication of such an effect.
Using the \cite{davis1951, chandrasekhar} and \cite{hildebrand} methods, we estimated the strength of the plane-of-the-sky field
and the ratio of turbulent-to-ordered field components in two regions of the cloud: one containing diffuse 
striations, and the other harbouring the highest column density filament. Our results indicate that the magnetic 
field is dynamically important in both regions. 
Combining our results, we find that differences of $6^\circ-30^\circ$ in the magnetic field inclination 
between two cloud regions can explain the observed polarization fraction differences. 
This difference can also be explained by a difference in the line-of-sight dispersion of the field of 
$10^\circ-25^\circ$.
Finally, we find that the polarization angles of the 
North Star \citep[]{heiles2000} and of RoboPol data in the surrounding area favour the scenario of the cloud being 
in front of the star.

\section*{Acknowledgements}

The authors thank L. Cambr\'esy for providing the extinction map and S. Clark and E. Koch for helpful 
advice on their codes.
We also thank D. Blinov, J. Liodakis, R. Skalidis, A. Tritsis and E. Palaiologou for their help throughout 
the duration of this project. We are grateful to M. Houde for his comments on the manuscript and to P. F. 
Goldsmith and D. Clemens for useful scientific discussions. We finally thank the anonymous reviewer for their insightful 
comments which helped significantly improve the manuscript.
G.V.P. and K.T. acknowledge support by FP7 through the Marie Curie Career Integration Grant 
PCIG-GA-2011-293531 ``SFOnset'' and partial support from the EU FP7 Grant PIRSES-GA-2012-31578 ``EuroCal''.
This research has used data from the \textit{Herschel} Gould Belt Survey (HGBS) project 
(http://gouldbelt-herschel.cea.fr). 
The HGBS is a \textit{Herschel} Key Programme jointly carried out by SPIRE Specialist Astronomy 
Group 3 (SAG 3), scientists of several institutes in the PACS Consortium (CEA Saclay, INAF-IFSI Rome and INAF-
Arcetri, KU Leuven, MPIA Heidelberg), and scientists of the \textit{Herschel} Science Center (HSC).

%%%%%%%%%%%%%%%%%%%%%%%%%%%%%%%%%%%%%%%%%%%%%%%%%%



%\bibliographystyle{mnras}

\begin{thebibliography}{99}
\bibitem[Andersson, Lazarian \& Vaillancourt (2015)]{andersson2015} Andersson B.-G., Lazarian A., Vaillancourt J. E., 2015, ARAA, 53, 501 
\bibitem[Andr\'e et al. (2014)]{andre2014} Andr\'e P., Di  Francesco  J.,  Ward-Thompson  D.,  Inutsuka S.-I., Pudritz R. E., Pineda J. E,  2014,  Protostars and Planets VI, 27
\bibitem[Andr\'e et al. (2010)]{andre2010} Andr\'e P. et al., 2010, A\&A, 518, L102
\bibitem[Arzoumanian et al. (2011)]{arzoumanian2011} Arzoumanian D. et al., 2011, A\&A, 529, L6A
\bibitem[Alves et al. (2008)]{alves2008} Alves F. O., Franco G. A. P., Girart J. M., 2008, A\&A, 486, 13A
\bibitem[Alves de Oliveira et al. (2014)]{deoliveira} Alves de Oliveira C., 2014, A\&A, 568, 98A
\bibitem[Barnes et al. (2015)]{barnes} Barnes P., Li D., Telesco C, Tanakul N., Marinas N., Wright C., Packham C., Pantin E., Roche P., Hough J., 2015, MNRAS, 453, 2622
\bibitem[Bensch et al. (2003)]{bensch} Bensch F., Leuenhagen U., Stutzki J., Schieder R., 2003, ApJ, 591, 1013
\bibitem[Bohlin et al. (1978)]{bohlin} Bohlin R. C., Savage B. D., Drake J. F., 1978, ApJ, 224, 132
\bibitem[Brunt et al. (2003)]{brunt2003} Brunt C. M., Heyer M. H., V\'azquez-Semadeni E., Pichardo B., 2003, ApJ, 595, 824B
\bibitem[Cambr\'esy et al. (2001)]{cambresy} Cambr\'esy L., Boulanger F., Lagache G., Stepnik B., 2001, A\&A, 375, 999
\bibitem[Chandrasekhar \& Fermi (1953)]{chandrasekhar} Chandrasekhar S., Fermi E., 1953, ApJ, 118, 113
\bibitem[Chapman et al. (2011)]{chapman2011} Chapman N. L., Goldsmith P. F., Pineda J. L., Clemens D. P., Li D., Kr\v{c}o M., 2011, ApJ, 741, 21C
\bibitem[Clark et al. (2014)]{clark2014} Clark S. E., Peek J. E. G., Putman M. E., 2014, ApJ, 789, 82
\bibitem[Crutcher (2004)]{crutcher2004} Crutcher R. M., Nutter D. J., Ward-Thompson D., Kirk J. M., 2004, ApJ, 600, 279
\bibitem[\protect\citeauthoryear{Dame, Hartmann \& Thaddeus}{2001}]{dame2001}
Dame T. M., Hartmann D., Thaddeus P., 2001, ApJ, 547, 792
\bibitem[Davis (1951)]{davis1951} Davis, Jr., L. 1951, Phys. Rev., 81, 890
\bibitem[Falceta-Gon\c{c}alves et al. (2008)]{falceta} Falceta-Gon\c{c}alves D., Lazarian A., Kowal G., 2008, ApJ, 679, 537
\bibitem[Fiege \& Pudritz (2002)]{fiege} Fiege J. D., Pudritz R. E., 2000, ApJ, 544, 830
\bibitem[Franco, Alves \& Girart (2010)]{franco2010} Franco G. A. P., Alves F. O., Girart J. M., 2010, ApJ, 723, 146
\bibitem[Franco \& Alves (2015)]{franco2015} Franco G. A. P., Alves F. O., 2015, ApJ, 807, 5
\bibitem[Gillmon \& Shull (2006)]{gillmon2006} Gillmon K., Shull J. M., 2006, ApJ, 636, 908
\bibitem[Girart, Rao \& Marrone (2006)]{girart} Girart J. M., Rao R., Marrone D. P., 2006, Science, 313, 5788, 812
\bibitem[Goldsmith et al. (2008)]{goldsmith2008} Goldsmith P. F. Heyer M., Narayanan G., Snell R., Li D., Brunt C., 2008, ApJ, 680, 428
\bibitem[Goldsmith (2013)]{goldsmith2013} Goldsmith P. F., 2013, ApJ, 774, 134
\bibitem[Goodman et al. (1990)]{goodman1990} Goodman A. A., Bastien P., Menard F., Myers P. C., 1990, ApJ, 359, 363
\bibitem[Gorbikov \& Brosch (2014)]{gorbikov} Gorbikov E., Brosch N., 2014, MNRAS, 443, 725
\bibitem[Greenberg (1968)]{greenberg} Greenberg J. M., 1968, Stars \& Stellar Systems, 7, 221
\bibitem[Grossmann et al. (1990)]{grossman} Grossmann V., Heithausen A., Meyerdierks H., Mebold U., 1990, A\&A, 240, 400
\bibitem[Hall (1955)]{hall1955} Hall J. S., 1955, Liege Symposium, 543
\bibitem[Heiles (2000)]{heiles2000} Heiles C., 2000, AJ, 119, 923H
\bibitem[\protect\citeauthoryear{Heithausen et al.}{1990}]{heithausen1990}
Heithausen A., Thaddeus P., 1990, ApJL, 353, L49
\bibitem[\protect\citeauthoryear{Heithausen et al.}{1993}]{heithausen1993} Heithausen A., Stacy J. G., de Vries H. W., Mebold U., Thaddeus P., 1993, A\&A, 268, 265
\bibitem[Heithausen (1999)]{heithausen1999} Heithausen A., 1999, A\&A, 349, L53
\bibitem[Heitsch et al. (2001)]{heitsch2001} Heitsch F., Zweibel E. G., Mac Low M.-M., Li P., Norman M. L., 2001, ApJ, 561, 800
%\bibitem[Hezareh et al. (2010)]{talayeh} Hezareh T., Houde M., McCoey C., Li H.-B., 2010, ApJ, 720, 603
\bibitem[Hildebrand et al. (2009)]{hildebrand} Hildebrand R. H., Kirby L., Dotson J. L., Houde M., Vaillancourt, J. E., 2009, ApJ, 696, 567
\bibitem[Hily-Blant \& Falgarone (2007)]{hily-blant2007} Hily-Blant P., Falgarone E., 2007, A\&A, 469, 173
\bibitem[Hily-Blant \& Falgarone (2009)]{hily-blant2009} Hily-Blant P., Falgarone E., 2009, A\&A, 500, L29
\bibitem[Houde et al. (2009)]{houde2009} Houde M., Vaillancourt J. E., Hildebrand R. H., Chitsazzadeh S., Kirby L., 2009, ApJ, 706, 1504
\bibitem[Keenan \& Babcock (1941)]{keenan} Keenan P. C., Babcock H. W., 1941, ApJ, 93, 64K
\bibitem[\protect\citeauthoryear{Koch \& Rosolowsky}{2015}]{koch2015} Koch E. W., Rosolowsky E. W., 2015, MNRAS, 452, 3435
\bibitem[Lee \& Draine (1985)]{leedraine} Lee H. M., Draine B. T. 1985, ApJ, 290, 211
\bibitem[Li et al. (2013)]{li2013} Li H.-B., Fang M., Henning T., Kainulainen J. 2013, MNRAS, 436, 3707
%\bibitem[Li \& Houde (2008)]{lihoude} Li H.-B., Houde M., 2008, ApJ, 677, 115
\bibitem[\protect\citeauthoryear{Malinen et al.}{2015}]{malinen2015} Malinen J. et al., 2015, MNRAS submitted, arXiv:1512.03775
\bibitem[Matthews (2001)]{matthews2001} Matthews B. C., Wilson C. D., Fiege J. D., 2001, ApJ, 562, 400
\bibitem[Men'shchikov et al.(2010)]{menshchikov} Men'shchikov A. et al.,2010, A\&A, 518, L103
\bibitem[Miville-Desch\^{e}nes et al. (2010)]{miville} Miville-Desch\^{e}nes M.-A. et al., 2010, A\&A, 518, L104
\bibitem[Mouschovias (1976)]{mouschovias1976} Mouschovias T. C., 1976, ApJ, 206, 753 
%\bibitem[Myers \& Goodman (1991)]{myers1991} Myers P. C., Goodman A. A., 1991, ApJ, 373, 509
\bibitem[Myers (2009)]{myers2009} Myers P. C., 2009, ApJ, 700, 1609 
\bibitem[Nakamura \& Li (2008)]{nakamura2008} Nakamura F., \& Li Z.-Y., 2008, ApJ, 687, 354
\bibitem[Ostriker,  Stone,  \& Gammie (2001)]{ostriker} 	Ostriker E. C., Stone J. M., Gammie C. F., 2001, ApJ, 546, 980
%\bibitem[Padoan \& Nordlund (2002)]{padoan2002} Padoan P., Nordlund \r{A}.  2002, ApJ, 576, 870
\bibitem[Padoan et al. (2001)]{padoan2001} Padoan P., Goodman A. A., Draine B. T., Juvela M., Nordland \r{A}. R, R\"{o}gnvaldsson O. E., 2001, ApJ, 559, 1005
\bibitem[Palmeirim et al. (2013)]{palmeirim} Palmeirim P. et al., 2013, A\&A, 550A, 38P
\bibitem[Panopoulou et al. (2014)]{panopoulou2014} Panopoulou G. V., Tassis K., Goldsmith P. F., Heyer M., 2014, MNRAS, 444, 2507P
\bibitem[\protect\citeauthoryear{Panopoulou et al.}{2015}]{panopoulou2015} Panopoulou G. V. et al., 2015, MNRAS, 452, 715
\bibitem[Pereyra \& Magalh\~{a}es (2004)]{pereyra2004} Pereyra A., Magalh\~{a}es A. M., 2004, ApJ, 603, 584
\bibitem[\protect\citeauthoryear{Pilbratt et al.}{2010}]{pilbratt2010} Pilbratt G. L. et al., 2010, A\&A, 518, L1
%\bibitem[Pillai et al. (2014)]{pillai} Pillai T., Kauffmann J., Tan J. C., Goldsmith P. F., Carey S. J., Menten K. M., 2015, ApJ, 799, 74P
\bibitem[Planck Collaboration Int. XIX (2015)]{planck19} Planck Collaboration et al., 2015, A\&A, 576, 104
\bibitem[Planck Collaboration Int. XX (2015)]{planck20} Planck Collaboration et al., 2015, A\&A, 576, 105 
\bibitem[Planck Collaboration Int. XXXII (2014)]{planck32} Planck Collaboration et al., 2014, A\&A, in press, arXiv:1409.6728
\bibitem[Planck Collaboration Int. XXXV (2015)]{planck35} Planck Collaboration et al., 2015, preprint, (arXiv:1502.04123)
%\bibitem[Savage et al. (1977)]{savage} Savage B. D., Bohlin R. C., Drake J. F., Budich W., 1977, ApJ, 216, 291
\bibitem[Savage \& Mathis (1979)]{savagereview} Savage B. D., Mathis J. S., 1979, ARA\&A, 17, 73S 
%\bibitem[Schlafly \& Finkbeiner (2011)]{schlafly2011} Schlafly E. F., Finkbeiner D. P., 2011, ApJ, 737, 103
\bibitem[Schlafly et al. (2014)]{schlafly2014} Schlafly E. F. et al., 2014, ApJ, 786, 29S
\bibitem[Schlegel et al. (1998)]{sfd} Schlegel D. J., Finkbeiner D. P., Davis M., 1998, ApJ, 500, 525
\bibitem[Schneider et al. (2013)]{schneider2013} Schneider N. et al., 2013, ApJL, 766, L17
\bibitem[Soler et al. (2013)]{soler2013} Soler J. D., Hennebelle P., Martin P. G. et al., 2013, ApJ, 774, 128
\bibitem[Sousbie (2011)]{sousbie} Sousbie T., 2011, MNRAS, 414, 350
\bibitem[Sugitani et al. (2011)]{sugitani2011} Sugitani K. et al., 2011, ApJ, 734, 63S
\bibitem[van den Bergh (1956)]{vandenbergh}van den Bergh, S., 1956, Z. Ap., 40, 249
\bibitem[Wagle et al. (2015)]{wagle} Wagle G. A., Troland T. H., Ferland G. J., Abel N. P., 2015, ApJ, 809, 17W
\bibitem[Ward-Thompson et al. (2010)]{WT2010} Ward-Thompson D. et al., 2010, A\&A, 518, L92
\bibitem[Zagury et al. (1999)]{zagury1999} Zagury F., Boulanger F., Banchet V., 1999, A\&A, 352, 645
\end{thebibliography}


%%%%%%%%%%%%%%%%%%%%%%%%%%%%%%%%%%%%%%%%%%%%%%%%%%

%%%%%%%%%%%%%%%%% APPENDICES %%%%%%%%%%%%%%%%%%%%%

%\appendix

%%%%%%%%%%%%%%%%%%%%%%%%%%%%%%%%%%%%%%%%%%%%%%%%%%

% Don't change these lines
\bsp	% typesetting comment
\label{lastpage}
\end{document}